\documentclass[journal]{IEEEtran}
\ifCLASSINFOpdf
   \usepackage[pdftex]{graphicx}
      
\else
\fi
\usepackage{amsmath}
\usepackage{mathtools}
\begin{document}
%
\title{Simple Fair Power Allocation for NOMA-Based Visible Light Communication Systems}
%
%
%


\author{Carlos Alberto Rodr\'{i}guez L\'{o}pez, and Vitalio~Alfonso~Reguera
\thanks{C. A. Rodr\'{i}guez L\'{o}pez is with the Central University of Las Villas e-mail: (crodrigz@uclv.edu.cu),}
\thanks{V. Alfonso Reguera is with the Federal University of Santa Maria e-mail: (vitalio.reguera@ufsm.br).}
}

\maketitle

\begin{abstract}
Non-orthogonal multiple access (NOMA) in the power-domain has been recognized as a promising  technique to overcome the bandwidth limitations of current visible light communication (VLC) systems. In this letter, we investigate the power allocation (PA) problem in an NOMA-VLC system under high signal-to-noise-ratio (SNR) regime. A simple fair power allocation strategy (SFPA) is proposed to ensure equitable allocation of transmission resources in a multi-user scenario. SFPA requires minimal channel state information (CSI), making it less prone to channel estimation errors. Results show that NOMA with SFPA provides fairer and higher achievable rates per user (up to 79.5\% higher in the studied setup), without significantly compromising the overall system performance.
\end{abstract}


%
\IEEEpeerreviewmaketitle
\section{Introduction}
In recent years non-orthogonal multiple access (NOMA) has drawn attention of many researchers as a promising technique to overcome the bandwidth limitations of current visible light communication (VLC) systems \cite{Obeed2019}. Under NOMA, users are multiplexed in the power-domain on the transmitting side and multi-user signal separation on the received side is performed based on successive interference cancellation (SIC) \cite{Islam2017}. Transmission in VLC systems is mostly based on the principle of intensity modulation of existing lighting sources, typically light-emitting diodes (LEDs). At the communication end-points, users are equipped with photodiodes (PDs) that perform the direct detection of the modulated optical signal. NOMA is considered a suitable alternative for VLC downlinks (NOMA-VLC), since power resources are least scarce in these networks, as compared to the bandwidth requirements of traditional orthogonal multiple access (OMA) mechanisms. Also, several studies have shown the superiority of NOMA over OMA in specific VLC scenarios \cite{Marshoud2016, Yin2016, Zhang2017, Yang2017, Fu2018}. 

An important issue in NOMA is to implement an efficient power allocation (PA) strategy. Currently, most frequently employed PA strategies in NOMA-VLC are the fixed power allocation (FPA) \cite{Lin2017}, and the gain ratio power allocation (GRPA) \cite{Marshoud2016}. With FPA the power is allocate without a comprehensive knowledge of channel state information (CSI). In this case, it is sufficient gather gross channel gain information to order the users in increasing order of signal strength. By using GRPA, the sum rate of the system can be improved by adapting the PA coefficients according to the estimated CSI. An enhanced PA strategy, named normalized gain difference power allocation (NGDPA) was proposed in \cite{Chen2018}. The performance of NGPDA was evaluated in an indoor 2×2 MIMO-NOMA-based multi-user VLC system, showing a significant sum rate gain over GRPA for large number of users. Both GRPA and NGDPA are more susceptible to CSI estimation errors than its FPA counterpart. Also, none of these PA strategies offers guarantees that the transmission capacity will be fairly distributed among VLC network users.

Fair resources allocation in NOMA-VLC has been studied in \cite{Yang2017, Fu2018, Liu2020}, from the perspective of an optimization problem. In \cite{Yang2017} the average and peak optical power constraints were considered, and an optimal power control algorithm was proposed using the Karush-Kuhn-Tucker (KKT) conditions for optimality. In \cite{Fu2018}, the optimal power ratio for an arbitrary number of multiplexed users in OFDM-NOMA-VLC was derived. The enhanced power allocation (EPA) method proposed in \cite{Fu2018} takes into account, along with detailed CSI feedback, the minimum requirement of signal-to-interference-plus-noise ratio (SINR) for each user. While in \cite{Liu2020}, the optimal PA was found by transforming the non-convexity of the original formulated model into a convex programming problem. Obviously, with these approaches, finding the optimal solution comes at the expense of a non-negligible increase in computational complexity.

In this letter, a low-complexity PA strategy that ensures a fair distribution of transmission capacity is proposed. The simple fair power allocation (SFPA) strategy presupposes a high transmission signal-to-noise-ratio (SNR), which is a common scenario to many practical VLC applications \cite{Obeed2019, Yin2016}. SFPA computations just involve the precise knowledge of the CSI for the user with best channel condition; which makes it less prone to channel estimation errors than other strategies that need to be aware of the CSI of the whole system. Numerical simulation results show that the proposed strategy exhibits higher transmission rates per user, when compared to PA strategies of similar computational complexity; without significantly compromising the overall system performance.  

\section{System Model}
\label{sec:System}

\begin{figure}[!t]
\centering
\includegraphics[width=0.8\columnwidth]{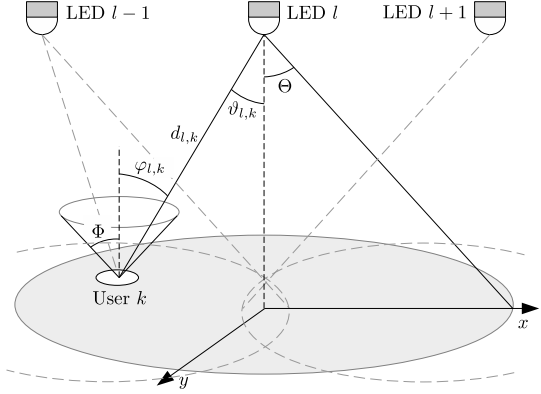}
\caption{LoS VLC downlink channel model}
\vspace{-20pt}
\label{fig:System}
\end{figure}

We consider an NOMA-VLC downlink with $L$ cooperating LEDs serving $K$ users randomly distributed in the coverage area. The users receive the line-of-sight (LoS) optical signal according to the channel model depicted in Fig. \ref{fig:System}. Assuming a Lambertian emission model, the channel gain between the LED $l$ and user $k$, is given by

\begin{equation}\label{eq:Lambertianmodel}
h_{l.k} = \frac{\mathcal{R} A_{p} (m+1)}{2 \pi d_{l,k}^{2}}\cos^{m}(\vartheta_{l,k})T(\varphi_{l,k})g(\varphi_{l,k})\cos(\varphi_{l,k}),
\end{equation}

\noindent
where $A_{p}$ is the physical area of the PD, $\mathcal{R}$ is the responsivity of the PD, $d_{l,k}$ is the distance between LED $l$ and user $k$, $m$ is the Lambertian emission order given by $m = \frac{-1}{\log_2(\cos(\Theta))}$, with $\Theta$ being the half-intensity radiation angle, $\vartheta_{l,k}$ and $\varphi_{l,k}$ are the LED irradiance angle and the PD incidence angle, respectively,  $T(\varphi_{l,k})$ represents the gain of the optical filter, and $g(\varphi_{l,k})$ is optical concentrator gain given by

\begin{equation}\label{eq:Concentratorgain}
g(\varphi_{l,k}) =
\begin{dcases}
\frac{n^2}{\sin^2 (\Phi)}, & 0 \leq \varphi_{l,k} \leq \Phi; \\
0, & \varphi_{l,k} > \Phi,
\end{dcases}
\end{equation}

\noindent
where $n$ is the refractive index and $\Phi$ is the semi-angle of the field-of-view (FoV).

Let $\rho_{l,i}$ denote the electrical power coefficient corresponding to the $i$th power-domain superposed user in LED $l$, satisfying $\sum_{i=1}^{K} \rho_{l,i} = 1$, then the received signal at user $k$ is

\begin{equation}\label{eq:recsignal}
y_k = P_o \sum_{l=1}^{L} h_{l,k} \sum_{i=1}^{K} \sqrt{\rho_{l,i}} \; x_{l,i} + z_k,
\end{equation}

\noindent
where $P_o$ is the LED output optical power, $x_{l,i}$ is the unitary optical power message signal intended for the $i$th user of LED $l$, and $z_k$ is additive white Gaussian noise of zero mean and power $\sigma_{k}^{2}$. Therefore, the total power, received at user $k$, can be expressed as

\begin{equation}\label{eq:signapower}
\vert y_k \vert^2= P_o^2 \Vert h_{k} \Vert^2 + \sigma_{k}^{2},
\end{equation}

\noindent
where $\Vert h_{k} \Vert^2 = \sum_{l=1}^{L} \vert h_{l,k} \vert^2$; without loss of generality, hereinafter we assume $\Vert h_{K} \Vert \geq \Vert h_{K-1} \Vert \geq \ldots \geq \Vert h_{1} \Vert$.  

Let $\alpha_i$ denote the fraction of the received total power allocated to the $i$th user. According to NOMA, users with worst channel condition, $\Vert h \Vert$, will receive a greater power fraction, i.e. $\alpha_{1} \geq \alpha_{2} \geq \ldots \geq \alpha_{K}$. Also, the SIC decoding follows the same order, decoding first user 1, then user 2 and so on up to decode user $K$. For a detailed description of superposition coding and SIC operation in NOMA, please see \cite{Islam2017}. For simplicity of analysis, we assume that perfect SIC can be performed without signal error propagation \cite{Yin2016,Chen2018}. Consequently the SINR experienced by user $k$ is

\begin{equation}\label{eq:sinr}
\text{SINR}_k = \frac{\alpha_k P_o^2 \Vert h_{k} \Vert^2}{(1 - \sum_{i = 1}^{k} \alpha_i) P_o^2 \Vert h_{k} \Vert^2 +\sigma_{k}^{2}}.
\end{equation}

It should be noted that the user power allocation per LED, $\rho_{l,i}$, must be carried out in such a way that the fraction of received power intended for the $i$th user, $\alpha_i$, conforms to a common SIC order. In other words, the SIC order should not be established on an independent per LED  basis, but following a single order for all common associated users, otherwise there would be an insurmountable ambiguity that would make decoding impractical. Considering the users that simultaneously receive transmissions from an LED set $G$, the combine superposition coding with common SIC decoding order can be simply implemented by making $\rho_{\forall l \in G ,i} = \alpha_i$. 

\section{Proposed power allocation}
\label{sec:PowerA}
With FPA $\alpha_{k} = \mu^{k}/\sum_{i=1}^{K} \mu^{i}$, where $\mu$ is a constant over the interval $[0,1)$, which is deliberately chosen to meet the desired system performance (e.g. maximize the sum rate). The appropriate selection of $\mu$ depends on the specific channel conditions, and therefore the FPA strategy does not adapt well to system changes (e.g. repositioning of users or variation in the number of serviced users). In what follows we propose a simple method to estimate the value of $\mu$ to ensure a fair resources distribution under high transmission SNR regime, which is a typical condition in VLC applications.
\subsection{Simple fair power allocation (SFPA) for high SNR}
Let $(1 - \sum_{i = 1}^{k} \alpha_i) P_o^2 \Vert h_{k} \Vert^2 \gg \sigma_{k}^{2}$ (i.e. interfering power from higher SIC order users be much greater than noise power), for $k<K$. Then, equation (\ref{eq:sinr}) can be approximated as

\begin{equation}
\label{eq:sinrhigh}
\text{SINR}_k = 
\begin{dcases}
\frac{\alpha_k}{1 - \sum_{i = 1}^{k} \alpha_i}, & 1 \leq k < K; \\[6pt]
\frac{\alpha_k P_o^2 \Vert h_{k} \Vert^2}{\sigma_{k}^{2}}, & k = K.
\end{dcases}
\end{equation}

Noticing that $\sum_{i=1}^{K} \mu^{i} = \mu (1 - \mu^K)/(1- \mu)$, for the FPA strategy, $\alpha_k$, can be expressed as

\begin{equation}\label{eq:alpha}
\alpha_k = \frac{\mu^{k-1}(1- \mu)}{1 - \mu^K}.
\end{equation}

Substituting (\ref{eq:alpha}) in the upper case of (\ref{eq:sinrhigh}) we get

\begin{equation}\label{eq:sinrw}
\text{SINR}_k =  \frac{1-\mu}{\mu(1 - \mu^{K-k})}, \quad 1 \leq k < K.
\end{equation}

From where, it can be observed that the SINR grows slightly with $k$ (i.e. with the improvement of channel conditions), and that for relatively small $\mu$, with increasing number of uses, $\frac{1-\mu}{\mu}$ becomes a tight lower bound of the SINR perceived by the weaker user. Then, in order to provide a fair allocation of resources, let us make the $\text{SINR}_K$ (i.e. the SINR for the user with the strongest channel conditions) be related to the SINR lower bound of as

\begin{equation}\label{eq:sinrbound}
\text{SINR}_K =  \left( \frac{1-\mu}{\mu} \right)^\beta,
\end{equation}

\noindent
where $\beta \geq 1$ is conveniently set to provide an excess rate for user $K$. That is, to control the proportion of rate allocated to the strongest user over the weakest user.  

Particularly, for $\beta = 1$, substituting (\ref{eq:sinrhigh}) and (\ref{eq:alpha}) in (\ref{eq:sinrbound}), after some algebraic work, we get

\begin{equation}\label{eq:mu}
\mu = \left( 1+ \gamma \Vert h_{K} \Vert^2 \right)^{-\frac{1}{K}},
\end{equation}

\noindent
where $\gamma = P_o^2 / \sigma_K^2$ is the transmission SNR, which is considered sufficiently high for the above results to hold. For $\beta > 1$, a numerical solution can be found at the expense of consuming more time and computing resources. 

Aimed to keep our proposal as simple as possible, SFPA is implemented by computing $\mu$ according to (\ref{eq:mu}) and then using (\ref{eq:alpha}) to obtain the PA coefficients. Thus, based on the previous reasoning, we expect to have a fair distribution of the transmission capacity among all competing users with minimum computational complexity. Furthermore, it is worth noting that for $\mu$ to be computed, along with the number of users, only the exact knowledge of the strongest user (user $K$) channel condition is required.  

\subsection{Fairness index}
Let $\bar{R_i}$ be the average rate for the $i$th user, ordered according to the instantaneous achievable rates. For the purpose of measuring fairness we use Jain's equation \cite{Jain1984}, which in this case can be expressed as 

\begin{equation}\label{eq:findex}
f(\bar{R}) = \frac{\left( \sum_{i=i}^{K} \bar{R_i} \right)^2}{K \sum_{i=i}^{K} \bar{R_i}^2}.
\end{equation}

This index measures the fairness with which the transmission capacity is allocated to users. If the value of the fairness index is 1, it indicates that all users receive the same average rate. If the transmission capacity is monopolized by few users, this index decreases near to 0.

\section{Simulation Results and Discussion}
\label{sec:Simulation}
In this section, we evaluate the performance of five different power allocation strategies through numerical simulation. We consider a $3 \times 3$ $\text{m}^2$ room, with 2 cooperating LEDs using NOMA with a common constructed superposition coding and SIC decoding order. Users were randomly positioning with vertical orientation. The setup and specific parameters adopted are summarized in Table \ref{tab:parameters}. It should be noted that in this scenario the coverage areas of the LEDs overlap at every possible position. Therefore, a high overall SNR is attained. Achievable rates were estimate using the tight capacity lower bound derived in \cite{Wang2013} (Eq. $(37)$, pp. 3775). For EPA a minimum required SINR of $1$ dB was considered in all the simulations. After completing all the iterations with random user distribution, the average rate perceived by the users and the average sum rate of the system were computed.

\begin{table}
\caption{Simulation Parameters}\label{tab:parameters}
\centering
\begin{tabular}{l|r}
\hline
Parameter name, notation & value\\    \hline
\hline
Number of LEDs, $L$ & $2$\\
Half-intensity radiation angle, $\Theta$ & $60^o$\\
LEDs' optical power, $P_o$ & $10$ W \\
Signal bandwidth, & $20$ MHz\\
Noise porwer spectral density, & $10^{-21} \, \text{A}^2/\text{Hz}$\\
Physical area of PDs, $A_p$ & $1 \, \text{cm}^2$ \\
FoV semi-angle of PDs, $\Phi$ & $60^o$\\
Refractive index, $n$ & $1.5$\\
Gain of the optical filter, $T(\varphi_{l,k})$ & 1\\
Responsivity of the PDs, $\mathcal{R}$ & $0.53$ A/W\\
Room horizontal dimensions, & $3 \times 3$ \\
LEDs' horizontal coordinates, $(x,y)$  & $(1,1), (2,2)$ \\
Vertical separation between LEDs and PDs, & $2$ m\\
Number of users, $K$ & $2-8$\\
Number of random user positions, & $10000$ \\ \hline
\end{tabular}
\vspace{-14pt}
\end{table}

\begin{figure}[!t]
\centering
\includegraphics[width=\columnwidth]{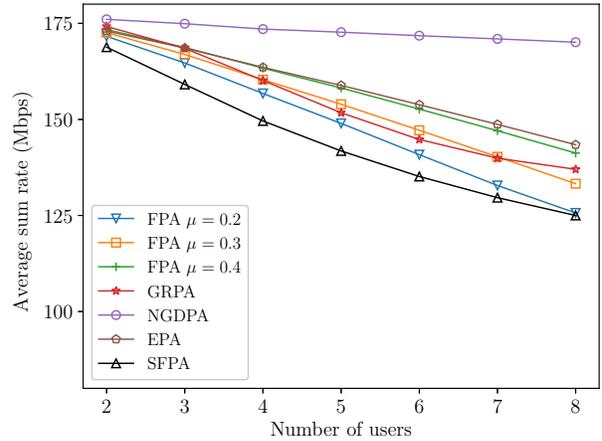}
\caption{Average sum rate for five different PA strategies using NOMA with an increasing number users} 
\label{fig:Sumrate}
\vspace{-10pt}
\end{figure}

Fig. \ref{fig:Sumrate} shows the average sum rate achieved using strategies FPA, GRPA \cite{Marshoud2016}, NGDPA \cite{Chen2018}, EAP \cite{Fu2018}, and SFPA. In this scenario, NGDPA outperforms all other strategies in terms of average sum rate, with SFPA exhibiting the lowest score. When compared to the optimal strategy with fairness support EPA, the decrease in the sum rate observed with SFPA is $12.8$\% in the worst case ($K = 8$). As demonstrated below, the decrease in overall system performance of SFPA is offset by a much fairer distribution of transmission capacity.

\begin{figure}[!t]
\centering
\includegraphics[width=\columnwidth]{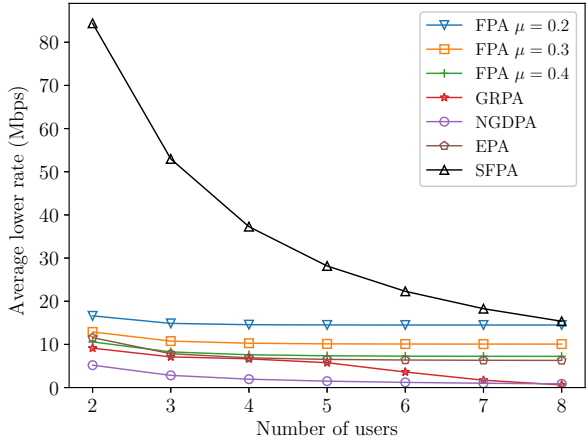}
\caption{Minimum rate perceived by users using NOMA with different PA strategies in a two LED scenario with high SNR} 
\label{fig:Minrate}
\vspace{-10pt}
\end{figure}

The results in Fig. \ref{fig:Sumrate} can be better understood by observing the average rate perceived by the least favored user (i.e. the user with the lowest rate) depicted in Fig. \ref{fig:Minrate}. As has been demonstrated in \cite{Chen2017}, the optimal PA strategy to maximizes the sum rate is the one that allocates the excess power to the strongest user, after having met the minimum rate constrains. Actually, this is the strategy followed by EAP. Moreover, it is obvious that without minimum rate constraint, the maximum sum rate is reached by allocating all the resources to the user with better channel conditions. As evidenced from Fig. \ref{fig:Minrate}, the outstanding sum rate achieved by NGDPA is at the expense of providing poor transmission capacity to less favored users. Among the other strategies, the difference in the average sum rate is less pronounced. Particularly for the SFPA strategy, its slightly lower aggregate performance (as compared to FPA, GRPA and EAP) is generously compensated by the significant increase of achievable lower rate. For instance, with exactly two users, the highest achieved average sum rate is $177$ Mbps with NGDPA, and the average rate for the least favored user, in the best of cases (FPA with $\mu = 0.2$), is $17$ Mbps. While with SFPA the average rate perceived per user is approximately $83$ Mbps (a $79.5$\% increment) yielding an average sum rate of $168$ Mbps (a $5.1$\% decrement).


\begin{figure}[!t]
\centering
\includegraphics[width=\columnwidth]{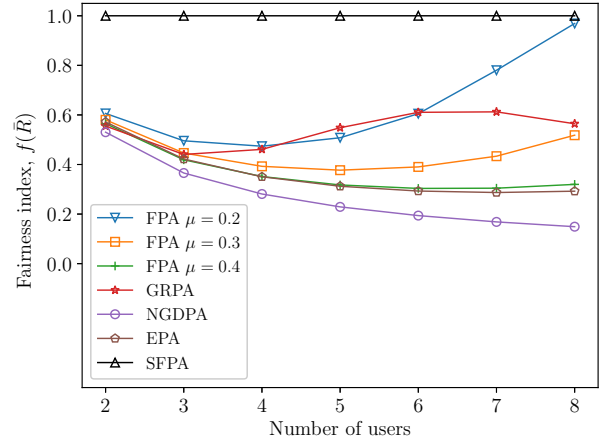}
\caption{Fairness index using NOMA with five different PA strategies} 
\label{fig:Fairness}
\end{figure}

The fairness index, computed using (\ref{eq:findex}), for the five PA strategies analyzed, is shown in Fig. \ref{fig:Fairness}. As expected, the ranking is inverted in relation to the one shown for the average sum rate, with SFPA being the strategy that offers the greatest equity in the distribution of transmission capacity. What is more, its fairness index is very close to 1 (i.e. all users experience almost the same transmission rate), regardless of the number of connected users. It is important to emphasize that most of the preceding works, aimed to provided a simple PA strategy, focus on maximizing the sum rate. However, as demonstrated above, this may be at the expense of a highly unfair distribution of transmission capacity. Conversely, the proposed SFPA strategy is capable of offering an acceptable overall performance, with comparable low complexity, while guaranteeing a fair resources allocation. 

\section{Conclusion}
In this letter, a novel PA strategy for NOMA-VLC systems was proposed. Unlike other low-computational complexity methods, the proposed strategy allows for a fair resources distribution, regardless of the number of competing users. The proposed SFPA strategy presupposes a high transmission SNR regime, which is a typical scenario for VLC applications. Also, differently from strategies that take advantage of full knowledge of channel conditions, the computation SFPA only involves the CIS of the stronger user, which makes it less prone to estimation errors. Results, for the studied setup, showed that it is possible to share the transmission capacity equitably without significantly compromising the overall system performance. This makes the SFPA strategy a suitable candidate to ensure high (up to $79.5$\% higher in our experiment) and fair transmission rates in future multi-user NOMA-VLC systems.  

\ifCLASSOPTIONcaptionsoff
  \newpage
\fi



\bibliographystyle{IEEEtran}
\bibliography{bare_jrnl}
\end{document}